\documentclass[12pt,a4]{article}   
\usepackage{epsfig} 
\usepackage{cite}

\newcommand{\Rbold}{\mbox{\boldmath$R$}}
\newcommand{\rbold}{\mbox{\boldmath$r$}}
\newcommand{\OmegaB}{\mbox{\boldmath$\Omega$}}

\begin{document}  

\thispagestyle{empty} 

\setlength{\baselineskip}{2.6ex} 

\begin{flushright}
Preprint PSI-PR-99-32 \\
December 15, 1999
\end{flushright}
\vspace*{10mm}

\begin{center}

{\Large\sf Collisional Quenching of the $2S$ State \\ of Muonic Hydrogen}
\\[5mm]
{\sc T.S. Jensen$^{\rm a,b}$ and V.E. Markushin$^{\rm a}$
\\[5mm]
{\em $^a$Paul Scherrer Institut, CH-5232 Villigen PSI, Switzerland}
\\[2mm]
{\em $^b$Institut f{\"u}r Theoretische Physik der Universit{\"a}t Z{\"u}rich,}\\
{\em  Winterthurerstrasse 190, CH-8057 Z{\"u}rich, Switzerland} 
}
\\[10mm]
\end{center} 


\begin{abstract}
We have calculated differential, total and transport cross sections for 
muonic hydrogen in the $n=2$ state  scattering from hydrogen. 
The metastable fraction of the $2S$ state that slows down below the 
$2P$ threshold without undergoing collisional quenching has been calculated as 
a function of the initial kinetic energy using a Monte Carlo kinetics program. 
Contrary to earlier estimates, the metastable fraction 
in the kinetic energy range of $2 - 5\;$eV cannot be neglected.   
\end{abstract} 

\clearpage

\section{Introduction}

The $2S$ state of muonic hydrogen offers interesting possibilities to do precision 
tests of QED and to determine the proton RMS charge radius 
(see \cite{ta} and references therein). An isolated $(\mu p)_{2S}$ is metastable  
with a lifetime mainly determined by muon decay (about $2.2\;\mu$s). In 
liquid or gaseous hydrogen the lifetime 
of the $2S$ state is shortened considerably because of Stark mixing followed by 
$2P\rightarrow 1S$  radiative transitions. 
If a sizeable fraction of muonic hydrogen atoms ends up in the $2S$ state  
with a sufficiently long lifetime, then  precision laser experiments with this 
metastable $2S$ state become feasible. 
 If the $(\mu p)_{2S}$ has kinetic energy below the $2P$ threshold 
(laboratory kinetic energy $T_0=0.3$~eV), then Stark transitions 
$2S\rightarrow 2P$ are energetically forbidden\footnote{Some quenching will, 
however, occur because $2S-2P$ mixing during collisions allows radiative 
transitions to the $1S$ state (See Refs.~\cite{mu,cb,mp}).}.     
The metastable fraction of $(\mu p)_{2S}$ in hydrogen depends on the 
kinetic energy at the time of formation.

 The first estimate of the $(\mu p)_{2S}$ lifetime was done by  
Kodosky and Leon \cite{kl}. They calculated the inelastic
$2S\rightarrow 2P$ cross section  in a semiclassical framework and 
concluded that the $2S$ state for $T>T_0$ will be rapidly depopulated except for 
very small  target densities. However, this model did not consider deceleration due
to elastic $2S\rightarrow 2S$ scattering.
A more elaborate approach was developed by Carboni and Fiorentini  \cite{cf}. They calculated
both elastic $2S\rightarrow 2S$ and inelastic $2S\rightarrow 2P$  cross sections quantum
 mechanically and estimated the probability for a $(\mu p)_{2S}$ atom
to slow down below threshold from a given initial energy.
The results of their calculations show that a sizeable fraction of $(\mu p)_{2S}$ 
formed at kinetic energies
less than 1.3~eV can slow down below the $2P$ threshold.

The metastable fraction of $(\mu p)_{2S}$ per stopped muon can in principle
be calculated in a cascade model which takes the different processes 
(Stark mixing, radiative decays, etc.) into account~\cite{ma1,ma2}. 
However, if one knows the fraction of stopped muons which reaches the $2S$ state
(regardless of energy) and the kinetic energy distribution on arrival in this
state, then it is sufficient to treat the final part of the cascade ($n=1,2$).
This information can be obtained  from experiments. The fraction of stopped muons which 
arrives in  the $2S$ state can be determined from the radiative yields~\cite{an2,br,la}: 
it was found in Ref.~\cite{an2} that between  2\% and 7\% of the $\mu p$ reach 
the $2S$ state in the pressure range $0.33-800$~hPa.   
The kinetic energy distribution for $\mu p$ in the $1S$ state, which for low pressures 
is expected to be very similar to that of the $2S$ state just after arrival, can be 
obtained from diffusion experiments~\cite{ha,ko}. 
The median energy is found to be about 1.5~eV for a target pressure of 
0.25~hPa \cite{ko}.

The purpose of this paper is to calculate the fraction 
of $\mu p$ in the $2S$ state which reaches kinetic energies below the $2P$ 
threshold as a function of the initial kinetic energy $T$.
We will also present a fully quantum mechanical calculation of $\mu p+$H 
differential cross sections which are used in our  Monte Carlo simulation of the kinetics.


The paper is organized as follows. 
The theoretical framework of the quantum mechanical calculation 
of the cross sections is outlined in Section~2. The calculated cross sections are 
discussed in Section~3. Section~4 presents the calculations of the metastable $2S$ 
fraction. The summary of the results is given in Section~5. 

Unless otherwise stated, atomic units ($\hbar=a_0=m_e=1$) are used throughout 
this paper. The unit of cross section is $a_0^2=2.8\cdot10^{-17} {\rm cm}^2 $.

\section{Quantum Mechanical Approach to the Calculation of the Cross Sections
 $(\mu p)_{nl}+{\rm H}\rightarrow (\mu p)_{nl^\prime}+{\rm H}$}

\begin{figure}
\begin{center}
\mbox{\epsfysize=3cm\epsffile{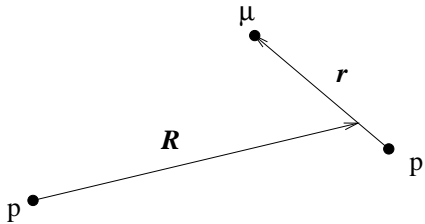}}
\end{center}
\caption{\label{FigCoord}%
Coordinates used in the calculations: $\Rbold$ is the
vector from the target proton to the center of mass of
the $\mu p$, $\rbold$ is the relative vector of the  $\mu p$ system.}
\end{figure}

For the benefit of the reader, we briefly describe the quantum mechanical calculation
of $\mu p+$H scattering in the coupled-channel approximation.  
The three-body wave function $\psi(\rbold,\Rbold)$ where the 
coordinates $\rbold$ and $\Rbold$ are defined in Fig.~\ref{FigCoord} satisfies 
the Schr\"odinger equation 
\begin{equation}
  H \psi(\rbold,\Rbold) = E \psi(\rbold,\Rbold)
\label{schroed1}
\end{equation}
where the Hamiltonian is given by
\begin{equation}
 H = -\frac{\nabla ^2}{2\mu} + H_{\mu p} + V(\rbold,\Rbold) \quad . 
\end{equation}
Here $\mu=m_pm_{\mu p}/(m_p+m_{\mu p})$
is the reduced mass of the $p-\mu p$ system, with $m_p$ being 
the proton mass and $m_{\mu p}$ the total $\mu p$ mass. 
The two-body Hamiltonian of the $\mu p$ atom, $H_{\mu p}$,  
includes the Coulomb interaction and a term that describes the shift of  
the $nS$ state (mainly because of the vacuum polarization) with respect to 
the states with $l>0$.  
For the case $n=2$ considered below, the $2S$ state is lower than the $2P$ 
by  $\Delta E=0.21$~eV.  
The much smaller fine and hyperfine structure splitting is neglected. 
The potential $V(\rbold, \Rbold )$ describes the interaction of the $\mu p$ 
system with the target proton\footnote{For the sake of simplicity, 
we ignore the fact that the protons are identical particles.}:  
\begin{equation}
  V(\rbold ,\Rbold  ) =
  \frac{1}{| \Rbold -\epsilon\rbold |} - 
  \frac{1}{|\Rbold  +(1-\epsilon)\rbold |}
\label{v}
\end{equation}
where $\epsilon=m_{\mu}/m_{\mu p}=0.101$.

Equation~(\ref{schroed1}) is solved in the coupled-channel approximation by using  a finite number of
basis functions to describe the state of the $\mu p$. For the problem of
 $nlm\rightarrow nl^\prime m^\prime$ scattering considered
in this paper, the  set of $n^2$ eigenstates  with principal
quantum number $n$ has been selected but the basis
 can be extended in a straightforward manner.
With $n$ fixed, let $\chi_{lm}(\rbold)$ denote the normalized eigenfunctions
of the atomic Hamiltonian $H_{\mu p}$ with the energy $E_{nl}$, 
the square of the $\mu p$ internal angular momentum
${\bf l}^2$ (eigenvalue $l(l+1)$) and its projection along the $z$-axis $l_z$ 
(eigenvalue $m$). The total wave function $\psi (\rbold,\Rbold)$ is expanded  as
follows
\begin{equation}
\psi (\rbold,\Rbold)
   =R^{-1}\sum_{JMLl}\xi_{JLl}(R){\cal Y}_{Ll}^{JM}(\OmegaB ,\rbold)
  \label{coupled}
\end{equation}
where 
\begin{equation}
 {\cal Y}^{JM}_{Ll}(\OmegaB ,\rbold) = 
 \sum_{M_Lm}\langle LlM_Lm|JM\rangle Y_{LM_L}(\OmegaB)\chi_{lm}(\rbold)
 , \quad \OmegaB = \Rbold/R  \quad . 
\label{caly}
\end{equation}
are simultaneous eigenfunctions of 
 ${\bf J}^2$, ${\bf L}^2$, ${\bf l}^2$ and $J_z$ with eigenvalues
 $J(J+1)$, $L(L+1)$, $l(l+1)$ and $M$, respectively. 
Here {\bf L} is the $p-\mu p$ relative
angular momentum,  ${\bf J}={\bf L}+{\bf l}$ is 
the total orbital angular momentum of the system.
 For a given value of $J$ the system of radial Schr\"odinger equations 
has the form
\begin{equation}
 \Big(-\frac{1}{2\mu}\frac{d^2}{dR^2}+\frac{L(L+1)}{2\mu R^2}+E_{nl}-E\Big)\xi_{JLl}(R)+
 \sum_{L^\prime l^\prime} \langle L^\prime  l^\prime JM |V|LlJM\rangle 
 \xi_{JL^\prime l^\prime }(R)=0
\label{schroed2}
\end{equation}
where the potential matrix elements are calculated in the basis (\ref{caly}):
\begin{equation}
 \langle \OmegaB,\rbold|LlJM\rangle ={\cal Y}^{JM}_{Ll}(\OmegaB,\rbold) \quad . 
\end{equation} 
The matrix elements of the potential~(\ref{v}) have been  calculated analytically; 
the corresponding formulas are rather lengthy and will be given elsewhere. From the asymptotic
form of the solution of the $n^2$ coupled\footnote{ 
Because of parity conservation the equations  decouple into two sets of respectively
$n(n+1)/2$ and $n(n-1)/2$ coupled equations. } equations~(\ref{schroed2}), the  scattering matrix $S$
is extracted and cross sections can be calculated using standard formulas. The scattering
amplitude for $nlm\rightarrow nl^\prime m^\prime$ is given by
\begin{equation}
f_{nlm\rightarrow nl^\prime m^\prime }(\OmegaB)=\frac{4\pi}{2i\sqrt{k^\prime k}}
\sum_{L^\prime L M_{L}^\prime}i^{L-L^\prime}Y_{L^\prime M_L^\prime}(\OmegaB )
\langle L^\prime  l^\prime M_L^\prime m^\prime |S-1|Ll0m\rangle Y_{L0}^{*}(0)
\end{equation}
where 
\begin{equation}
\langle L^\prime l^\prime  M_L^\prime m^\prime |S|Ll0m\rangle =\sum_{JM}
\langle L^\prime  l^\prime M_L^\prime  m^\prime |JM\rangle 
  \langle JM|Ll0m\rangle \langle L^\prime l^\prime J|S|LlJ\rangle \quad . 
\end{equation}
As a consequence of rotational symmetry, 
the matrix elements $\langle L^\prime l^\prime J|S|LlJ\rangle $ do not depend on the quantum number  
$M$. The differential cross sections for the transitions $nl\rightarrow nl^\prime $ are given by
\begin{equation}
\frac{d\sigma_{nl\rightarrow nl^\prime}}{d\Omega}=\frac{1}{(2l+1)}
 \sum_{m^\prime m}\frac{k^\prime}{k}|f_{nlm\rightarrow nl^\prime m^\prime }|^2
\end{equation} 
where $k$ and $k^\prime$ are the magnitudes of the  
relative momenta in the initial and final state, correspondingly.

The total cross sections of the transitions $nl\rightarrow nl^\prime$ have  the form
\begin{equation}
\sigma_{nl\rightarrow nl^\prime}=\frac{1}{(2l+1)}\frac{\pi}{k^2}\sum_{J}(2J+1)
  \sum_{LL^\prime}|\langle L^\prime l^\prime J|S-1|LlJ\rangle |^2
\end{equation}
and the corresponding transport cross sections are  given by 
\begin{equation}
  \sigma^{tr}_{nl\rightarrow nl^\prime }=\int d\Omega (1-\cos\theta)
  \frac{d\sigma_{nl\rightarrow nl^\prime}}{d\Omega} \quad . 
\end{equation}

  In order to treat the long distance behaviour of the $\mu p+$H interaction 
properly, the effect of electron screening must be taken into account.  
This is done by multiplying the matrix elements 
$\langle L^\prime  l^\prime JM |V|LlJM\rangle$ 
in  Eq.~(\ref{schroed2}) by the screening factor 
\begin{equation}
 F(R) = (1+2R+2R^2) e^{-2R}
\end{equation}
which corresponds to the assumption that the electron of the hydrogen atom 
remains unaffected in the $1S$ state during the collision.

For  $p-\mu p$ separations $R$ smaller than a few units of the $\mu p$ Bohr 
radius, $a_\mu=0.0054$, our model cannot be expected to be valid because
the truncated set of basis functions in Eq.~(\ref{coupled}) 
is not sufficient to describe the total 
three-body wave function $\psi(\rbold,\Rbold)$.
Furthermore, exchange symmetry between the two protons must be taken into account.
We can estimate the sensitivity of our results to the short range part of 
the interaction by using the dipole approximation for the potential~(\ref{v}). 
The interaction in the dipole approximation is given by the first nonzero 
term in the expansion of  Eq.~(\ref{v}) in inverse powers of $R$: 
\begin{equation}
  V_{DA}(\rbold,\Rbold) = \frac{\rbold\cdot\Rbold}{R^3}=\frac{z}{R^2} \quad . 
\label{vda}
\end{equation}
A certain problem arises in the dipole approximation for a few low partial waves 
($J\leq 5$): the Schr{\"o}dinger equation
becomes ill defined because of the attractive $1/R^2$ 
singularity\footnote{This is a problem only in the 
dipole approximation. The exact matrix elements are all finite for  $R=0$.}.
Following Ref.~\cite{cf} we cure this difficulty by placing an infinitely 
repulsive sphere of radius $r_{\mathrm{min}}$ around the target proton. The sensitivity 
of the results to this cutoff parameter $r_{\mathrm{min}}$ will be used below as an 
estimate of the importance of detailed description of the interaction 
at short distances.    

In this paper we are interested in the $2l\rightarrow 2l^\prime$ transitions,
and only four states $n=2$ are used to describe the $\mu p$ part of the 
total wave function.  The four coupled second order 
equations~(\ref{schroed2}) are solved numerically for  $J=0,1,...,J_{\mathrm{max}}$ 
where the highest partial wave $J_{\mathrm{max}}$ is chosen large enough to 
ensure the convergence of the partial wave expansion at given collision 
energy.  
  
Until now we have considered the $\mu p$ collisions with the atomic target. 
Treating the collisions with hydrogen molecules is a formidable task (even for 
$\mu p$ in the ground state~ \cite{adam}) which we do not attempt here.  
The inelastic threshold $T_0$ for $2S\rightarrow 2P$ transitions is 0.44~eV for 
atomic target and 0.33~eV for a molecular target. To get the correct threshold 
value for the inelastic cross sections one can substitute the atomic hydrogen 
mass with the molecular one. 
By varying $r_{\mathrm{\mathrm{\mathrm{min}}}}$ and the target mass one can 
obtain an estimate of the theoretical uncertainty of our approach.

The present model for calculating cross sections for $(\mu p)_{n=2}-$H scattering 
is a straightforward extension of the one by Carboni and Fiorentini~ \cite{cf}. 
There are three major differences:
we solve the four coupled differential equations exactly while Ref.~\cite{cf}
treated non adiabatic terms as a perturbation. The second difference is that
we include the full angular coupling while Ref.~\cite{cf} omitted some minor terms. Finally,
we use exact matrix elements for the $\mu p-p$ interaction while 
Ref.~\cite{cf} considers only the dipole approximation.
These  approximations were justified by  Ref.~\cite{cf}  
as follows: for small kinetic energies the velocity of the muonic hydrogen atom is so
low that the motion can be regarded as nearly adiabatic. 
The angular coupling terms that were 
omitted in Ref.~\cite{cf} are of the order 1 which is much smaller than the remaining
angular coupling terms of the order of $J(J+1)$ (the angular momenta  
as high as $J \sim 15$ contribute to the $2S\rightarrow 2P$ cross section at $T=1$~eV). 
The electric field from a hydrogen atom
is strong enough to induce Stark transitions in the $\mu p$ for the distances  
$R \sim a_0$. Therefore, the regions where the dipole approximation is valid 
($R\gg a_\mu$) are supposed to be most important. 
The diffusion experiments~\cite{ha,ko} have shown that a sizeable fraction of
$(\mu p)_{n=2}$ atoms has kinetic energies of several eV. In  this high energy
region the non adiabatic couplings become strong and the model of Ref.~\cite{cf}
can not be expected to give accurate results.
  
The problem of $\mu p+$H scattering has  been treated fully quantum mechanically in 
Refs.~\cite{pp1,ppstark,ppdif,pptot}. However, these calculations did not include 
the $2S-2P$ energy splitting and the question of the metastability of $(\mu p)_{2S}$ 
was not addressed.
  Stark mixing has been studied in the semiclassical straight-line-trajectory 
approximation in Refs.~\cite{kl,lb,th}. We have calculated the $2S\rightarrow 2P$ 
Stark mixing cross sections in the semiclassical approach 
in order to compare with the quantum mechanical results. 
 A more detailed comparison between semiclassical and quantum mechanical 
calculations of $\mu p+{\rm H}$ scattering will be given elsewhere.

\section{The Cross Sections of $(\mu p)_{2l}$ Scattering from Hydrogen}

Using the method described in Section~2 the S-matrix has been calculated
for the laboratory kinetic energy range $T_0<T<6$~eV. 
Unless otherwise explicitly stated, the results shown are obtained with 
the exact potential~(\ref{v}). Electron screening is always taken into account.
Both atomic and molecular mass of the target 
($M_{\mathrm{target}}=M_{{\rm H}}$, $M_{{\rm H}_2}$) have been used. 
  
\begin{figure}
\begin{center}
\mbox{\epsfysize=6cm\epsffile{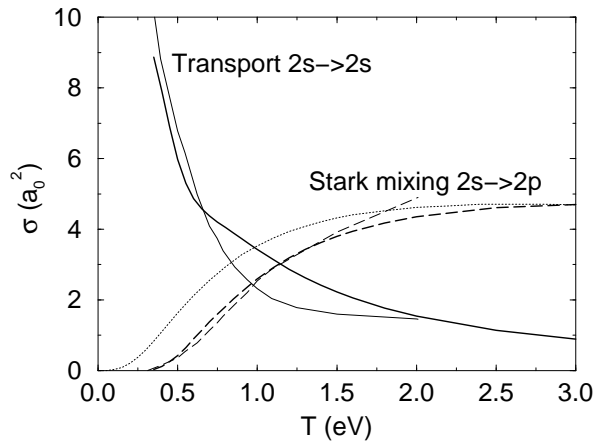}}
\end{center}\vspace*{-1.5\baselineskip}
\caption{\label{FigTrX} 
Transport cross section $2S\rightarrow 2S$ (solid lines) and Stark mixing cross 
section $2S\rightarrow 2P$ (long-dashed lines) vs. laboratory kinetic energy $T$. 
The thick lines are the results
of the present calculations, the thin lines are obtained from
Fig.~3 of Ref.~\cite{cf} and the dotted line shows the result of the semiclassical
calculation in the straight-line-trajectory approximation. The quantum mechanical
cross sections are calculated with $M_{\mathrm{target}}=M_{\rm H_2}$.}
\end{figure}

Figure~\ref{FigTrX} shows the  $2S\rightarrow2S$ transport cross section and the 
$2S\rightarrow2P$ Stark mixing cross section  in comparison 
with the results from Ref.~\cite{cf} (the molecular mass is used in both 
cases). 
There is a good agreement for the Stark mixing $2S\rightarrow 2P$ cross 
section below 1.7~eV. For the $2S\rightarrow 2S$ transport cross section, 
the agreement is fair, with the discrepancy being typically less than 30\%.  

In order to estimate the theoretical uncertainty of our approach, we calculated 
the cross sections in the dipole approximation with short distance cutoff 
$0.01\leq r_{\mathrm{min}} \leq 0.05$ for  
$M_{\mathrm{target}}=M_{\rm H}$ and $M_{\mathrm{target}}=M_{\rm H_2}$. 
For fixed  value of $M_{\mathrm{target}}$, the cross sections for the three reactions  
$2S\rightarrow 2P$, $2P\rightarrow 2S$ and  $2P\rightarrow 2P$ are weakly dependent 
on $r_{\mathrm{min}}$. This shows that these reactions are dominated by the 
long range part of the interaction $V(\rbold,\Rbold)$.  
The only process rather sensitive to the value of 
$r_{\mathrm{min}}$ is the elastic 
scattering $2S\rightarrow 2S$. This can be understood by considering the adiabatic 
energy curves for  low angular momentum. 
The energy curve which corresponds asymptotically to the $2S$ state is attractive 
while those corresponding to the three $2P$ states are repulsive. 
Therefore, in the adiabatic approximation the $2S\rightarrow 2S$ cross sections are 
expected to depend on the short range part of the potential while this is not 
the case for $2P\rightarrow 2P$ scattering. 
  At energies above 2 eV the semiclassical approximation is in a good agreement
with our quantum mechanical results.
 However, this semiclassical approximation  does not  treat
the threshold behaviour correctly.

\begin{figure}
\begin{center}
{\small \hspace{0.5cm} (a) \hspace{6cm} (b) }
\mbox{\epsfysize=5cm\epsffile{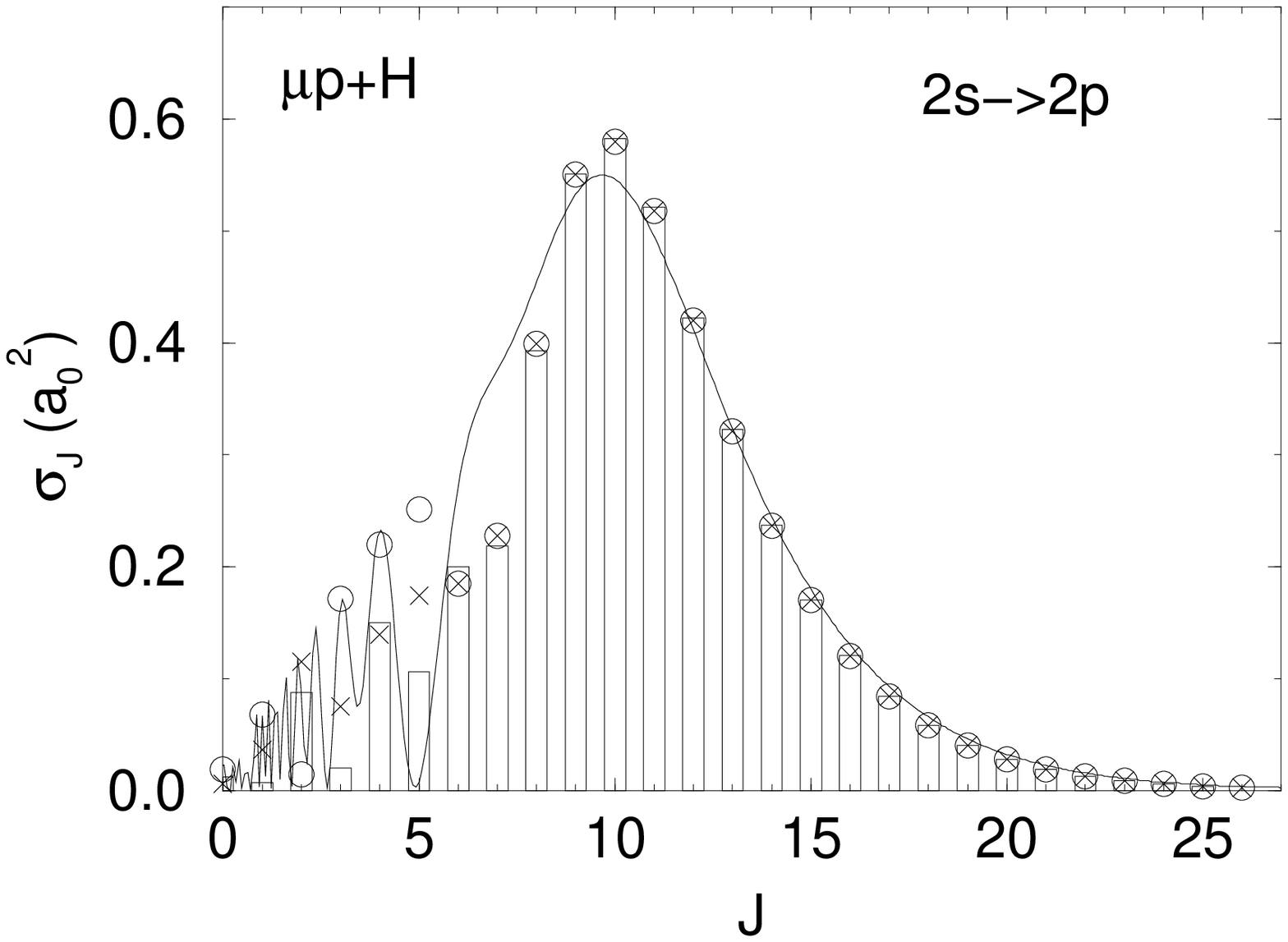}
      \epsfysize=5cm\epsffile{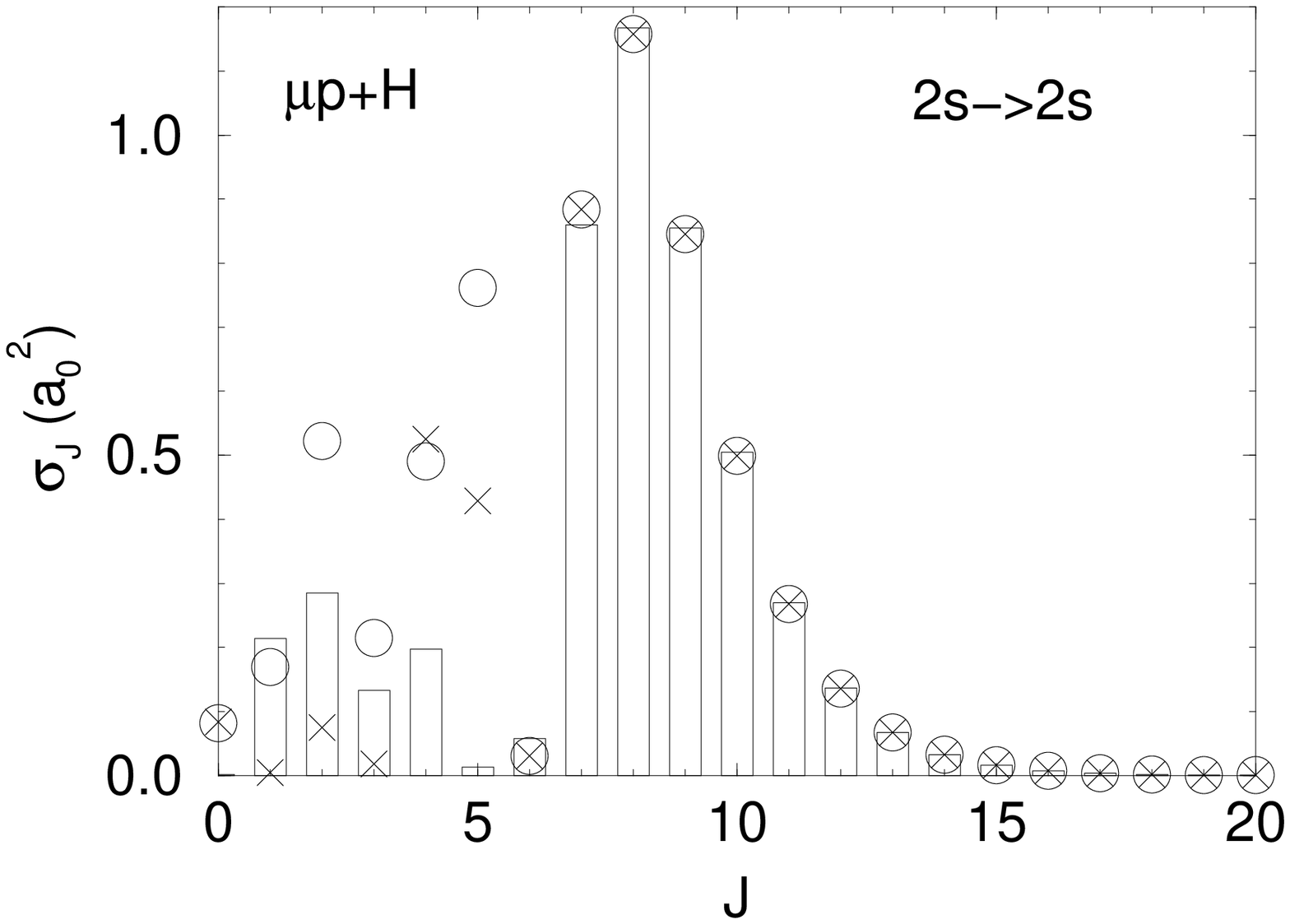}
}
\end{center}\vspace*{-1.5\baselineskip}
\caption{\label{FigPWX}%
Partial wave cross sections for the reactions  $2S\rightarrow 2P$
(a) and $2S\rightarrow 2S$ (b) at laboratory  kinetic energy  3~eV and 
$M_{\mathrm{target}}=M_{\rm H}$. 
The histograms show the results for  exact matrix elements; 
the  dipole approximation  with $r_{\mathrm{min}}=0.01$ and $r_{\mathrm{min}}=0.05$
is shown  by o and $\times$, respectively. The semiclassical
result for the  $2S\rightarrow 2P$ transition is shown with a solid line.
}
\end{figure}

A more detailed comparison of the approximations used can be done by plotting 
the $J$ dependence of the partial wave cross 
sections $\sigma_J$ at fixed energy as shown in Fig.~\ref{FigPWX}. 
The quantum mechanical results obtained for the exact
potential and the dipole approximation with the short range cutoff agree
well for angular momentum  $J>5$ while the lowest partial waves are sensitive to 
the short range behaviour of the approximating potentials.
The reason is that for $J>5$ the centrifugal barrier
is strong enough to prevent the  $\mu p$  from approaching close to the
target proton. For $J\leq 5$ the $\mu p$ can get very close
to the proton and the use of a small number of atomic orbitals is
not sufficient  ---  a better description in this region is needed in a true 
three--body framework. It is seen that a substantial  part of the 
$2S\rightarrow 2S$ cross section comes from partial waves with low $J$, 
so this  result also explains why the uncertainty of the  elastic $2S$  cross section 
is larger than  for  the other reactions. 
   The semiclassical calculation can be compared with  the partial wave cross
sections by using the relation between the impact parameter $\rho$, the relative
momentum $k$
and the angular momentum $J$
\begin{equation}
 k\rho=J+1/2  \quad . 
\end{equation}
For large $J$ (large impact parameter) there is a very good
agreement between the semiclassical contribution to the  $2S\rightarrow 2P$
cross section  and the quantum mechanical partial wave result.
   An example of the differential cross sections for the reaction $2S\rightarrow 2S$ given 
in Fig.~4 shows a characteristic pattern with a strong forward peak and 
a set of maxima and minima, which is in qualitative
agreement with Ref.~\cite{ppdif} where  the adiabatic approach was used.  

\begin{figure}
\begin{center}
{\small \hspace{1cm}(a) \hspace{6cm} (b)} \\
\mbox{
\mbox{\epsfysize=4.8cm \epsffile{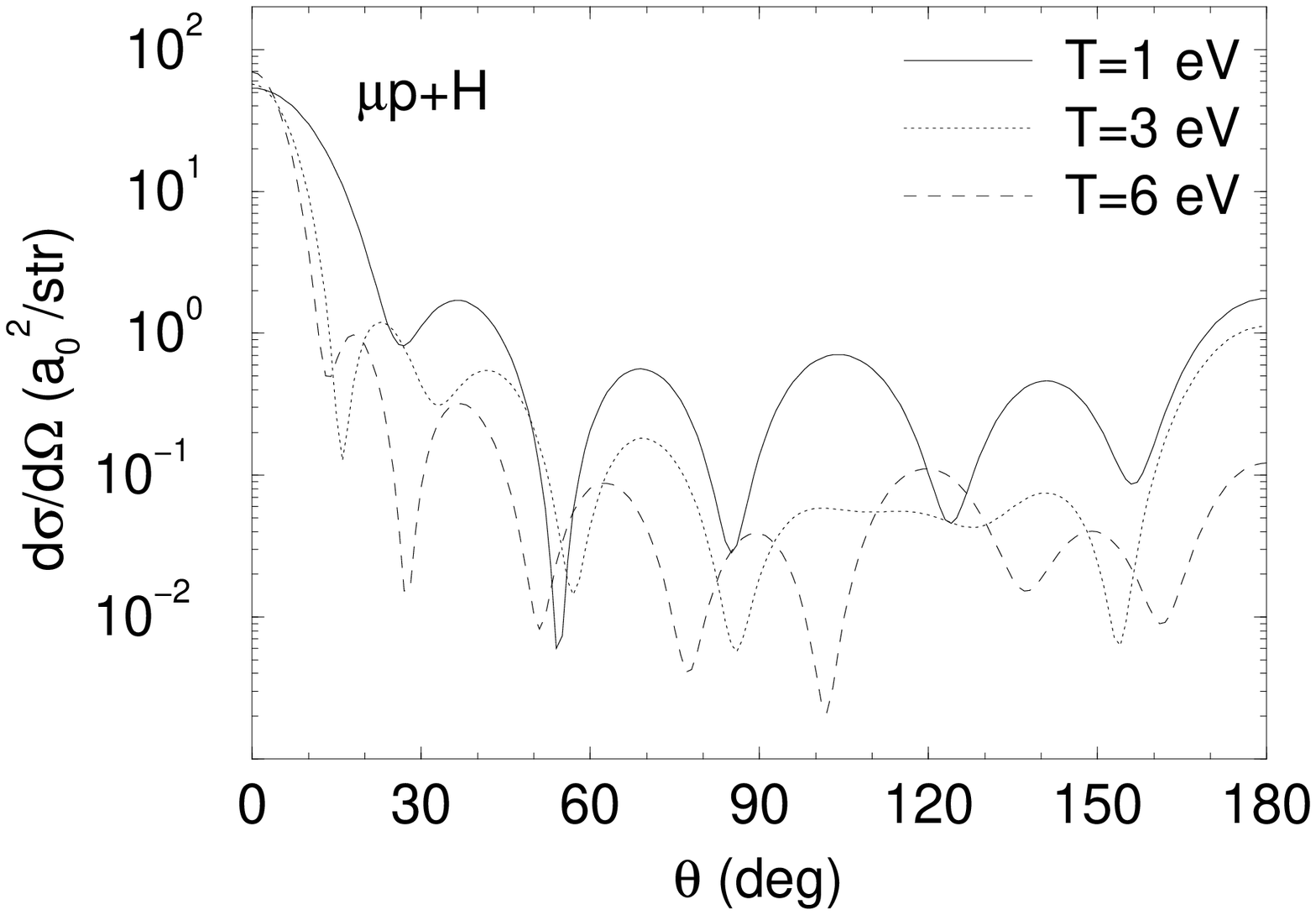}} 
\mbox{\epsfysize=4.8cm \epsffile{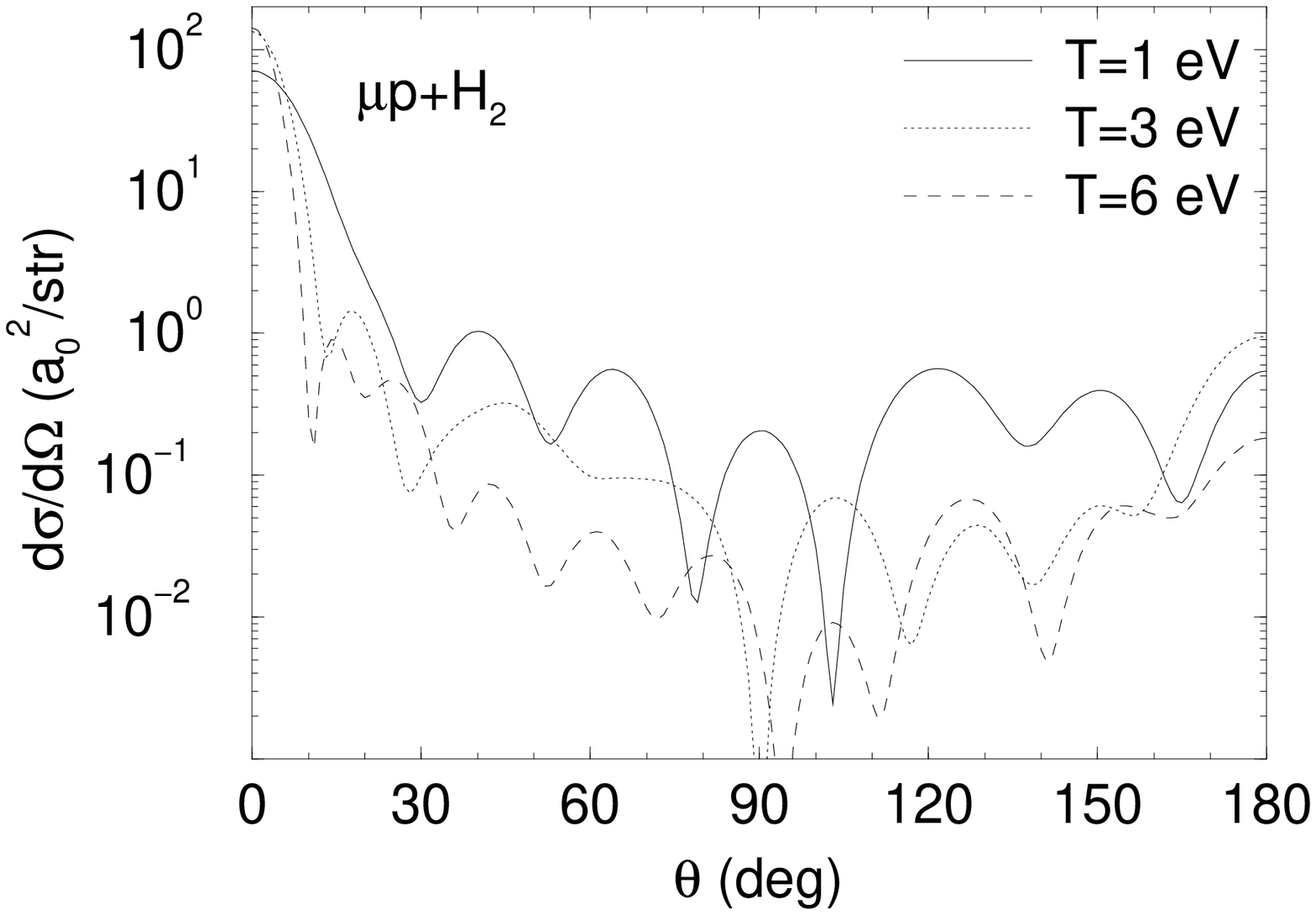}} 
}
\end{center}\vspace*{-1.5\baselineskip}
\caption{\label{FigDX}%
Differential $2S \rightarrow 2S$ cross sections vs. CMS scattering angle $\theta$ 
for three different laboratory kinetic energies: 
(a) $M_{\mathrm{target}}=M_{\rm H}$, (b) $M_{\mathrm{target}}=M_{{\rm H}_2}$. }
\end{figure}

\section{The Surviving Fraction of the Metastable $(\mu p)_{2S}$ State}

The surviving metastable  fraction $f(T)$ is defined as the probability
that the $\mu p$ atom in the $2S$ state with initial kinetic energy $T$ reaches 
the energy below the $2P$ threshold by slowing down in elastic collisions. 
Assuming that the rate of the radiative transition 
$2P\rightarrow 1S $, $\lambda_{2P\rightarrow 1S}=1.2\cdot 10^{11}~{\rm s}^{-1}$, 
is much larger than the Stark mixing rate\footnote{With our result for the Stark 
mixing rate $2P\rightarrow 2S$ at 1~eV as a function of the target density $N$, 
$ \lambda_{2P\rightarrow 2S}=Nv\sigma_{2P\rightarrow 2S}\approx4\cdot 10^{12}(N/N_0)~{\rm s}^{-1}$
where $N_{0}$ is the liquid hydrogen density $4.25\cdot 10^{22}$~atoms/cm$^3$, the range of
validity is $N\ll 0.03N_0$.}, the surviving fraction $f(T)$  was estimated
in \cite{cf} by the formula
\begin{equation}  
f(T)=\exp\left(-\frac{(m_{\mu p}+M_{\mathrm{target}})^2}{2m_{\mu p}M_{\mathrm{target}}}
 \int_{T_{0}}^{T}\frac{\sigma_{2S\rightarrow 2P}(T^\prime)}{T^\prime 
  \sigma_{2S\rightarrow 2S}^{tr}(T^\prime)}dT^\prime\right)
\label{formula}
\end{equation}
with $M_{\mathrm{target}}=M_{{\rm H_2}}$.
It was  found  that a sizeable fraction of $(\mu p)_{2S}$ atoms formed at 
kinetic energies below 1.3 eV slows down below threshold.

\begin{figure}
\begin{center}
\mbox{\epsfysize=6cm\epsffile{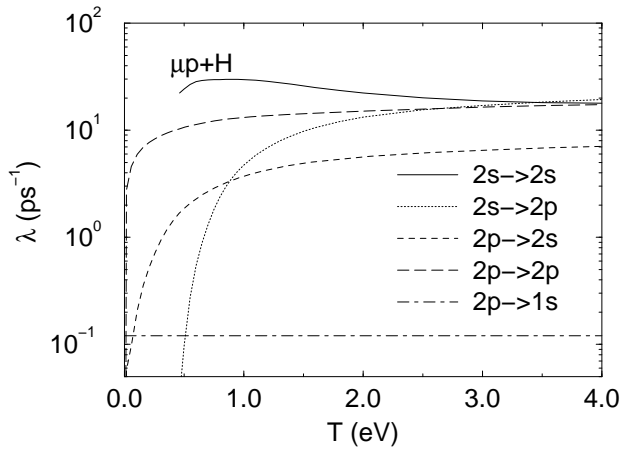}}
\end{center}\vspace*{-1.5\baselineskip}
\caption{\label{FigCR}
Collisional rates for $2l\rightarrow 2l^\prime$  in liquid hydrogen 
($N=N_0$, $M_{\mathrm{target}}=M_{\rm H}$) and the radiative $2P\rightarrow 1S$ 
transition rate.}
\end{figure}

Equation~(\ref{formula}) is based on the approximation of continuous energy loss. 
To provide a more realistic treatment
of  the evolution in kinetic energy we use a Monte Carlo program 
based on the differential cross sections for the four processes
$2S\rightarrow 2S$, $2S\rightarrow 2P$, $2P\rightarrow 2S$ and  $2P\rightarrow 2P$. 
In addition to the collisional 
processes, the  $2P\rightarrow 1S$ radiative transition is also included in the 
code. The fate of a $\mu p$ formed in the $2S$ state with kinetic energy $T$ 
is thus either to undergo $2P\rightarrow 1S$ radiative transition after the Stark mixing 
$2S\rightarrow 2P$ or to end up in the $2S$ state with kinetic energy 
below the threshold with probability $f(T)$.  
 Figure~\ref{FigCR} shows the rates, 
$\lambda_{2l\rightarrow 2l^\prime}=N_0v\sigma_{2l\rightarrow 2l^\prime}$, 
for the collisional transitions in liquid hydrogen in comparison with 
the radiative deexcitation rate $\lambda_{2P\to 1S}$. 
In liquid hydrogen the Stark mixing rates are so large that the $\mu p$ states are 
expected to be statistically populated for kinetic energies $T\ge 2\;$eV 
(where threshold effects can be neglected).   

\begin{figure}
\begin{center}
\mbox{\epsfysize=6cm\epsffile{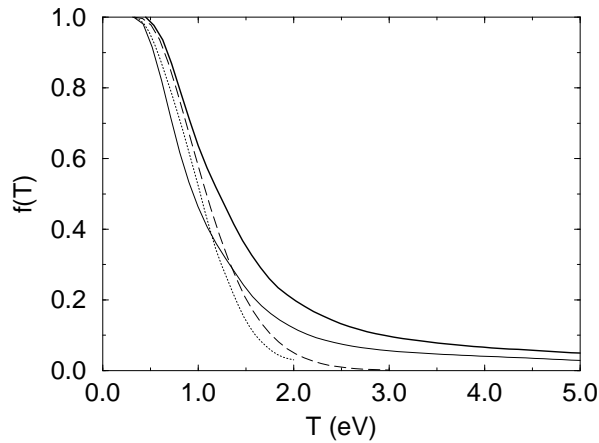}}
\end{center}\vspace*{-1.5\baselineskip}
\caption{\label{Figf} 
The metastable surviving fraction $f(T)$ of the $(\mu p)_{2S}$ states 
vs. initial kinetic energy $T$. 
The thick and thin solid lines show the results of 
Monte Carlo calculations with $M_{\mathrm{target}}=M_{\rm H}$ and  $M_{\mathrm{target}}=M_{\rm 
H_2}$, respectively. The dotted line shows the result of Carboni and
Fiorentini~ \cite{cf} and
the long-dashed line shows $f(T)$ calculated using the method of Ref.  \cite{cf}
(Eq.~(\ref{formula})) with our cross sections. The results are valid for densities  
$N\leq 10^{-2}N_0$.
}
\end{figure}

  Figure~\ref{Figf} shows the surviving fraction $f(T)$ calculated with 
the Monte Carlo program for target density $10^{-6} < N/N_0 < 10^{-2}$.  
The approximation~(\ref{formula}) gives somewhat higher values for  
the survival probability than the exact kinetics calculation at $T<1.4\;$eV.  
The Monte Carlo results at high energies ($T>1.5\;$eV) are significantly  
larger than those obtained from Eq.~(\ref{formula}) where continuous 
energy loss is assumed. The reason is that the backward scattering 
(see Fig.~4) with maximum possible energy loss plays an important role 
in bringing the $(\mu p)_{2S}$ atoms below the $2P$ 
threshold for higher energies.

 In order to estimate the theoretical uncertainty of $f(T)$ we performed 
the Monte Carlo calculation with the cross sections obtained in the dipole 
approximation.  In all cases the Monte Carlo calculations are consistent with the results 
corresponding to the exact potential: at 2 eV the surviving  fraction is in the range
$15-20\%$ for atomic target mass and $10-16\%$ for molecular target mass. 
  The use of the target mass $M_{\rm H}$ instead of $M_{\rm H_2}$ leads to somewhat
higher survival fractions because of a simple  kinematical reason: 
the loss of kinetic energy in a collision with the same angle in the CMS is larger 
for the target of smaller mass and, furthermore, 
the inelastic threshold is higher for the scattering from the atomic target. 
Merely substituting the atomic mass with the molecular mass for the hydrogen 
target does not account for 
the additional energy loss due to rotational and vibrational excitations
of H$_2$. One would therefore expect that the slowing down
process is more efficient than this model suggests and the survival probability  
calculated with molecular target mass is underestimated. The opposite is true for
calculations with atomic target: here the transfer of kinetic energy from 
the $\mu p$ to the 
individual hydrogen atoms is not restricted by molecular bindings. Thus results
with atomic target probably give somewhat optimistic results for the
surviving fraction.

\section{Conclusion}

The main results of this paper can be summarized as follows. The detailed Monte Carlo 
kinetics calculations predict the surviving metastable fraction of the $2S$ state of $\mu p$
to be larger than $50\%$ for the initial kinetic energy 1~eV in agreement with earlier 
estimates~ \cite{cf}. For higher initial kinetic energies, our result is significantly
larger than the earlier estimates: the surviving metastable fraction for $T=5$~eV is
about $4\%$. This effect is due to a sizeable contribution of backward scattering in
elastic collisions.

   Our Monte Carlo calculations are based on the cross sections 
calculated in the coupled-channel approximation. The main limitation of this method
for the problem concerned comes from the use of a small number of atomic states
to describe the $\mu p$ system and the neglect of the molecular structure of the target.
A more accurate treatment of the $\mu pp$
three body problem is needed in order to do reliable calculations for a few 
lowest partial wave amplitudes.
Our approach, however, is well suited for the description of the collisions 
with the characteristic scale of impact parameters of the order of $a_0$ 
which is exactly the case for the problem involved.  Therefore our results 
provide a significantly improved basis for a better estimate of the metastable 
$(\mu p)_{2S}$ fraction \cite{ExpR98-99} which is very important  
for the planned Lamb--shift experiment at PSI \cite{ta}.

Further details and more results concerning the scattering of the 
$\mu p$ atoms in the excited states $n \geq 2$ will be published elsewhere.

\section*{Acknowledgement}

We thank P.~Hauser, F.~Kottmann, L.~Simons, D.~Taqqu, and R.~Pohl for 
fruitful and stimulating discussions and M.P.~Locher for useful 
comments.   

\bibliographystyle{unsrt}

\end{document}